\documentclass[12pt,preprint]{aastex}

\usepackage{natbib}

\shorttitle{Cosmic-ray antiproton flux}
\shortauthors{Boezio et al.}

\newcommand{\ap}{\mbox{$\overline{{\rm p}}$ }}

\begin{document}

\title{The Cosmic-Ray antiproton flux between 3 and 49 GeV.}

\author{M. Boezio\altaffilmark{1}, V. Bonvicini,
P. Schiavon,  A. Vacchi, and N. Zampa}
\affil{ Dipartimento di Fisica dell'Universit\`{a} and Sezione INFN di 
Trieste, Via A. Valerio 2, I-34147 Trieste, Italy}

\author{D. Bergstr\"{o}m, P. Carlson, T. Francke, and S. Grinstein}
\affil{ Royal Institute of Technology (KTH),
           S-104 05 Stockholm, Sweden}

\author{M. Suffert}
\affil{ Centre des Recherches Nucl\'{e}aires, BP20,
F-67037 Strasbourg-Cedex, France}

\author{M. Hof, J. Kremer, W. Menn, and M. Simon}
\affil{ Universit\"{a}t Siegen, 57068 Siegen, Germany}

\author{S. A. Stephens}
\affil{ Tata Institute of Fundamental Research, Bombay 400 005, India}

\author{M. Ambriola, R. Bellotti, F. Cafagna, F. Ciacio, M. Circella, and C.
De Marzo}
\affil{ Dipartimento di Fisica dell'Universit\`{a} and Sezione INFN
di Bari, Via Amendola 173, I-70126 Bari, Italy}

\author{N. Finetti\altaffilmark{2},
P. Papini, S. Piccardi, P. Spillantini, E. Vannuccini}
\affil{ Dipartimento di Fisica dell'Universit\`{a} and Sezione INFN di Firenze,
Largo Enrico Fermi 2, I-50125 Firenze, Italy}

\author{S. Bartalucci, M. Ricci}
\affil{ Laboratori Nazionali INFN, Via Enrico Fermi 40, CP 13, I-00044
Frascati, Italy}

\author{M. Casolino, M.P. De Pascale, A. Morselli,  P. Picozza,
and R. Sparvoli}
\affil{ Dipartimento di Fisica dell'Universit\`{a} and Sezione INFN di Roma,
Tor Vergata, Via della Ricerca Scientifica 1, I-00133 Roma, Italy}

\author{J.W. Mitchell, J.F. Ormes, and R.E. Streitmatter}
\affil{ Code 661, NASA/Goddard Space Flight Center, Greenbelt, MD 20771, USA}

\and

\author{U. Bravar and S.J. Stochaj}
\affil{ Box 3-PAL, New Mexico State University, Las Cruces, NM 88003, USA}

\altaffiltext{1}{Also at Royal Institute of Technology (KTH),
           S-104 05 Stockholm, Sweden\\
           Electronic address: mirko.boezio@trieste.infn.it}
\altaffiltext{2}{Now at Dipartimento di Fisica dell'Universit\`{a} dell'Aquila,
Aquila, Italy}

\begin{abstract}
We report on a new measurement of the cosmic ray antiproton spectrum.
The data were collected by the balloon-borne experiment \mbox{CAPRICE98}, which
was flown on 28-29 May 1998 from
Fort Sumner, New Mexico, USA.
The experiment used the NMSU-WIZARD/CAPRICE98 balloon-borne magnet
spectrometer equip\-ped with a gas Ring Imaging Cherenkov (RICH)
detector, a time-of-flight system, a tracking device
consisting of drift chambers and a superconducting magnet
and a silicon-tungsten calorimeter. 
The RICH detector was the first ever flown capable of mass-resolving 
        charge-one particles at energies above 5 GeV.

A total of 31 antiprotons with rigidities 
between 4 and
50~GV at the spectrometer 
were identified with small backgrounds from
other particles. The absolute antiproton energy spectrum
was determined
in the kinetic energy region at the top of the atmosphere between 
3.2 and 49.1~GeV.
We found that the
observed antiproton
spectrum and the antiproton-to-proton ratio are consistent
with a pure secondary origin. However, a primary component may not 
be excluded.

\end{abstract}


\section{Introduction}
Detailed measurements of the cosmic-ray antiproton energy spectrum provide
important information concerning the origin and propagation 
of cosmic rays in the Galaxy. In fact,
antiprotons are a natural product of interactions between
cosmic rays and the interstellar matter. Moreover,
antiprotons can be produced
by exotic sources such as evaporation of primordial black holes
\citep{haw74,kir81,mak96} and annihilation of supersymmetric particles
\citep{ste85,bot98,ull99a,ull99b}.
The measurement of the antiproton spectrum at
energies above a few GeV permits the study of these topics free of 
uncertainties associated with
the secondary antiproton production such as nuclear sub-threshold effects
and of the uncertainties in the solar modulation effect. Furthermore,
it permits us to investigate the possible annihilation of heavy supersymmetric
particles \citep{ull99c}.

Several measurements of the cosmic-ray antiprotons have been performed since
the first detection by \citet{gol79}. However, most of these experiments have
measured the antiproton spectrum at energies below 4~GeV (see
\citet{ori00} and references therein). Only two experiments
\citep{gol79,hof96} have obtained
antiproton results at energies above 4~GeV and these 
results differ by a large amount.
We report in this paper 
a new observation of antiprotons with energies
up to 50~GeV obtained with the CAPRICE98
experiment. This apparatus was
launched by balloon from Fort Sumner, New Mexico (34.3$^{\circ}$ North
Latitude, 110.1$^{\circ}$ West Longitude) and landed
close to Holbroke, Arizona
(34.$^{\circ}$ North Latitude, 104.1$^{\circ}$ West Longitude),
on May 28 and 29, 1998, at an atmospheric
pressure of 4.2 to 6.2 mbar for 21 hours and average value of the vertical
cut-off rigidity of about 4.3~GV \citep{she83}.

Preliminary results on the antiproton to proton ratio
from CAPRICE98 were reported earlier \citep{ber00a}. Here,
we present the absolute energy
spectrum of antiprotons in the energy region at the top of the
atmosphere between 3 and 49~GeV.
We also describe in detail
the analysis of the flight data.
The detector system is described in Section 2,
the data analysis in Section 3 and the results are presented and discussed in
Section 4.

\section{The CAPRICE98 apparatus}
\begin{figure}
\plotone{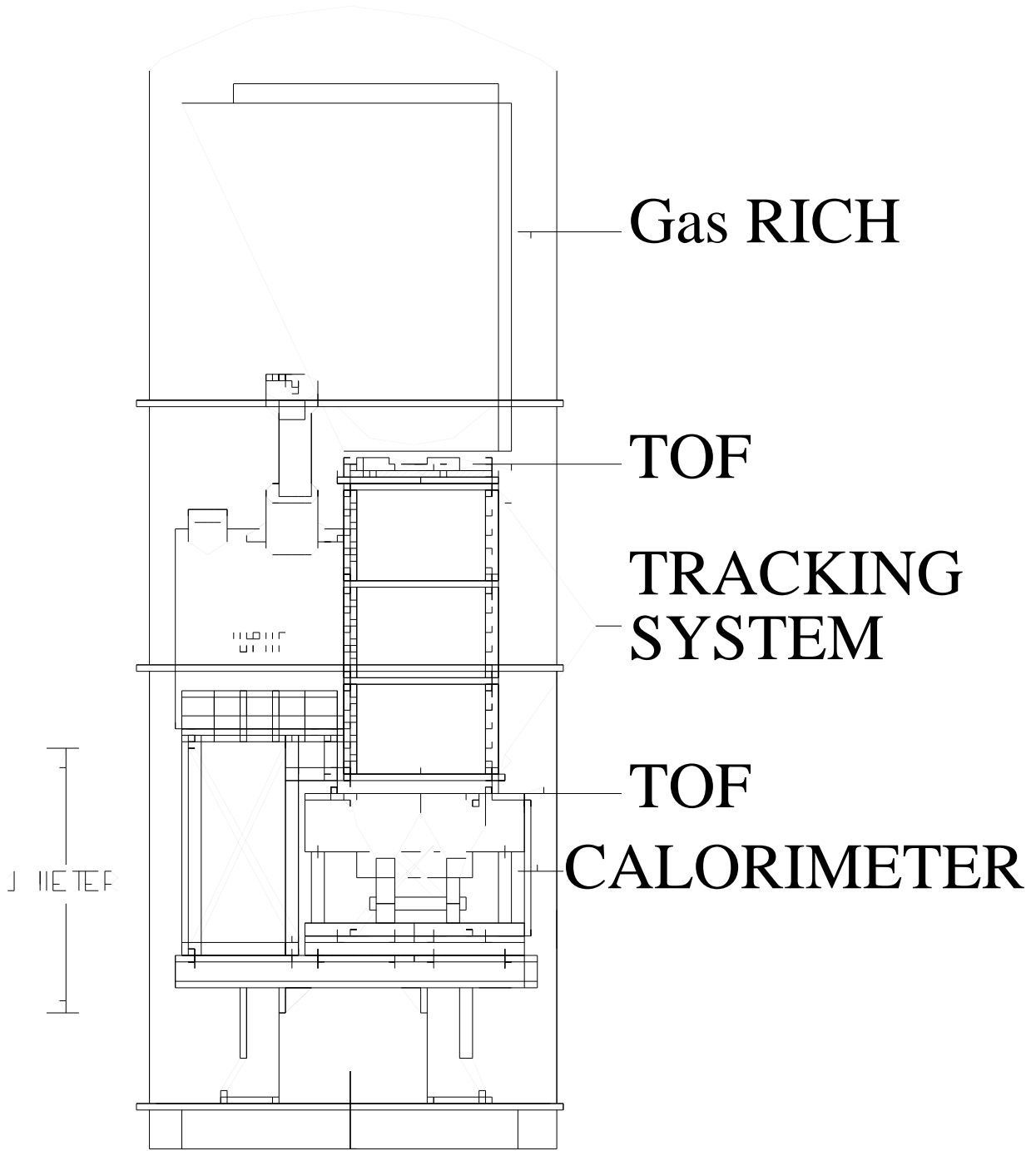}
\vspace*{1cm}
\figcaption[f1.eps]{Schematic view of the CAPRICE apparatus.
 \label{FigGon}}
\end{figure}

Figure~\ref{FigGon} shows the NMSU-WiZard/CAPRICE98 spectrometer
\citep{amb99}.
   It included from top to bottom: a gas Ring Imaging Cherenkov (RICH)
detector, a time-of-flight (ToF) system, a tracking system
consisting of drift chambers and a superconducting magnet
and a silicon-tungsten imaging calorimeter.

\subsection{The Gas RICH detector}
The Ring Imaging Cherenkov (RICH) detector
was designed primarily to identify
antiprotons in the cosmic rays in a large background of electrons, muons and
pions \citep{car94}.
The RICH detector \citep{fra99,ber00b}
consisted of a photosensitive multiwire proportional chamber (MWPC)
and a
1~m tall radiator box, filled with high purity C$_{4}$F$_{10}$ gas.
When a charged
 particle with $ \beta > 1/{\rm n}$, where n is the refractive index
and $\beta = v/c$ ($v$ being the particle velocity and $c$ the speed of
light), passed 
through the RICH detector, Cherenkov light was emitted in the gas
radiator along the trajectory. 
The Cherenkov light was emitted 
at an angle ($\theta_c$)
determined by the Cherenkov
relation ($cos(\theta_c) = \frac{1}{\beta \cdot {\rm n}}$),
and creating a cone of light in the same
 direction as the particle trajectory.

The cone of light 
after traversing the radiator volume was reflected
 back and focused by a spherical mirror toward
the MWPC. There the Cherenkov light interacted with a photosensitive
 gas, tetrakis-dimethyl-amino-ethylene (also called TMAE), and photoelectrons
 were produced. These electrons were amplified and 
then detected by induced pulses
in a matrix pad plane. This plane had 
an area of 51.2$\times$51.2~cm$^{2}$, divided in
$64 \times 64$ pads of size $8 \times 8$ mm$^2$,
where the cone of Cherenkov light gave a ring-like
image. The size of the ring was dependent on the velocity of the particle.
The ring diameter increased from 0 at
the RICH threshold (about 18~GV for protons) to
about 11~cm for a $\beta \simeq 1$ particle.
For $\beta \simeq 1$ charge one particles, an average of
12~photoelectrons per event were detected.

About half of the particles triggered by the 
instrument passed through the MWPC, where they ionized the gas. 
The ionization signals were 
amplified and detected by the pad plane along with the Cherenkov signals.

\subsection{The time-of-flight system}

The time-of-flight system consisted of two layers of plastic scintillators,
one placed above the tracking system and the other below, as
indicated in Figure~\ref{FigGon}.
Each layer was divided into two paddles with a size of 25$\times$50~cm$^{2}$
and a thickness of 1~cm. The material
used was Bicron 401. Each paddle had two
5~cm diameter photomultiplier tubes, one at each end.
The distance between the two scintillator layers was 1.2~m.

The signal from each photomultiplier was split in two parts, one was sent
to an analog-to-digital converter and the other to a time-to-digital
converter.
In this way, the time-of-flight system provided both energy loss (dE/dX) and
timing information.
The scintillator signals also provided the 
trigger for the data acquisition system.

\subsection{The tracking system}

The tracking system consisted of a superconducting magnet and three
drift chambers.
The average maximum detectable
rigidity (MDR) was 330~GV.

The magnet \citep{gol78} consisted of a single coil of 11,161
turns of copper-clad
   NbTi wire. The outer diameter of the coil was 61~cm and the inner diameter
36~cm. The coil was placed in a dewar filled with liquid helium
surrounded by a vacuum shell close in a
   second dewar filled with liquid nitrogen that reduced the rate of 
evaporation of
   liquid helium and enabled to attain a life time of about
100 hours for the superconducting magnet.
The operating current was
set at 120~A, producing an
inhomogeneous field of approximately 4~T at the center of the coil.

The three drift chambers \citep{hof94} used for the
trajectory measurements were
physically identical. The lateral sides
of the
chamber box were made from 1~cm thick epoxy-composite plates, while
the open top and bottom sides were covered with 160~$\mu$m thick
copper plated mylar windows. The inner gas volume of the box
was of size $47 \times 47 \times 35$~cm$^{3}$.
The drift chamber had six layers, each
layer containing sixteen 27.02~mm wide drift cells, for
measurements in the x-directions and four layers for the y-direction.
A high efficiency ($\approx 99\%$
for a single drift cell) and
an average spatial resolution better than 100~$\mu$m were found.
The three drift chambers
provided a total of 18 position measurements in the direction of maximum
bending, {\em x }direction, and 12 along the perpendicular view,
{\em y} direction.
Using the position information together with the map of the magnetic field,
the rigidity of the particle was determined. 

\subsection{The Calorimeter}

The silicon tungsten calorimeter flew in several
balloon-borne experiments. The CAPRICE98 configuration also was used in
the CAPRICE94
\citep{boe97} and CAPRICE97 \citep{kre99} experiments.
The calorimeter \citep{boc96,ric99}
was designed to distinguish non-interacting minimum ionizing particles,
hadronic cascades and electromagnetic showers.

The calorimeter consisted of eight 48 $\times$ 48~cm$^{2}$ silicon planes
interleaved with seven layers of tungsten converter, each one
radiation length (X$_{0}$) thick.
A single plane consisted of an array of 8$\times$8 pair of silicon
detectors.
Each detector had a total area of 60$\times$60~mm$^{2}$ and was divided into
16~strips, each of width 3.6~mm.
The detectors were mounted back-to-back with perpendicular strips to give
$x$ and $y$ readout.
The strips of each detector were daisy-chained longitudinally to form one
single strip 48~cm long.
Taking into account all the material, the calorimeter had a total
thickness of 7.2~X$_{0}$.

\section{Data analysis}
The analysis was based on 21 hours of data for a total acquisition
time of 67240 seconds under an average residual atmosphere of 5.5 g/cm$^{2}$.
The fractional live time during the
   flight was $0.4865 \pm 0.0002$ resulting in a total live time
($T_{live}$) of $32712 \pm 13$ s.

Antiprotons are a very rare component of the cosmic radiation. They
must be distinguished from a large background of protons and electrons.
Furthermore, products of interactions of cosmic rays in the 
atmosphere above
the payload, such as
muons
and pions, are a significant background for antiproton measurements
performed with
balloon-borne experiments.
For these reasons strict selection criteria had to be applied on 
the data acquired from each detector. The
rigidity range of the antiproton analysis was 4 to 50~GV. The lower limit was
due to the geomagnetic cut-off of the experiment, while the upper limit was
based on the RICH ability to reliably identify antiprotons from
other particles at maximum Cherenkov angle
($\beta \simeq 1$).
At 50~GV the (anti)proton Cherenkov angle became
less than 3 standard deviations away
from the Cherenkov angle of $\beta \simeq 1$ particles.

\subsection{Antiproton and proton selection}

\subsubsection{Tracking}

The primary
 task of the tracking system was to precisely measure 
the sign and absolute value of the deflection (1/rigidity) of the
 particle traversing the apparatus. Events with more
than one track, such as products of interactions, were eliminated.
For this reason a set of strict selection
 criteria was imposed on the quality of the fitted tracks. These criteria
 were based partly on experience gained during the analysis of data from a
 similar tracking system \citep{hof96,mit96,boe97}:
\begin{enumerate}
    \item At least 12 (out of 18) position measurements in the x direction
    (direction of maximum bending)
    and 8 (out of 12) in the y direction were used in the fit.
    \item There should be an acceptable chi-square for the
    fitted track in both directions with stronger requirements on the
    x-direction.
\end{enumerate}

In addition to these criteria, we required that the value of the
deflection, as determined using only the tracking information from
the top half of the spectrometer, be consistent with the value
determined using only the bottom half of the spectrometer. This
additional constraint was imposed on events below the RICH threshold
for antiprotons. For events where Cherenkov light was detected (above
18 GV) the particles could be distinguished using the calculated
mass. To remove the contamination from spillover protons in
the antiproton sample, a concern at high rigidities, we required
that the uncertainty in the deflection, estimated on an
event-by-event  basis, be less that 0.008~(GV)$^{-1}$ \citep{gol91}.
This value was chosen as a compromise between
   rejection power for spillover protons and efficiency of the condition in 
   the high rigidity range. It is worth pointing out that the Cherenkov angle 
   determined by the RICH detector provided an additional check 
on the deflection (see also section~\ref{s:spil}) with a comparable 
uncertainty at 50 GV.

\subsubsection{Scintillators and time-of-flight}

The information of the time-of-flight system, with a time resolution of
230~ps, was used to select downward moving particles.
The dE/dx information from the top scintillators was used to
reject alphas and heavier particles as well as
mul\-ti\-par\-ti\-cle
events coming from interactions above the top scintillator.
This was done by requiring the following two conditions.
\begin{enumerate}
    \item dE/dx losses in the top scintillator less than 1.8~mip
    (where a mip is the most probable energy loss
    for a minimum ionizing particle).
    \item Only one paddle hit in the top scintillator plane.
\end{enumerate}

Antiprotons interacting in the calorimeter could produce backscattered
particles that traverse the bottom scintillator paddles giving an additional
signal. None of these cases significantly affected the
performance of the tracking system and calorimeter.
Therefore, no restrictions were put
on the bottom scintillators.

\subsubsection{Calorimeter}
The calorimeter was primarily used to identify electrons.
   The longitudinal and transverse
   segmentation of the calorimeter combined with the measurement of the energy
   lost by the particle in each silicon strip resulted in a high
   identification power for electromagnetic showers.
In the analysis presented in this paper, the calorimeter was
   used to reject events
   with electromagnetic showers (see Bergstr\"{o}m
2000c\footnote{Ph.D. thesis 2000,
Royal Institute of Technology, is available at:
http://www.\-particle.\-kth.\-se/group\_docs/\-astro/\-research/\-references.html}
for a
description of the selection criteria), hence reducing the electron
contaminations in the antiproton sample. The procedure followed was
similar to the one used in the CAPRICE94 antiproton analysis \citep{boe97}.
The selection was designed to reject electrons while
   keeping as large an antiproton fraction as possible.

\begin{figure}
\plotone{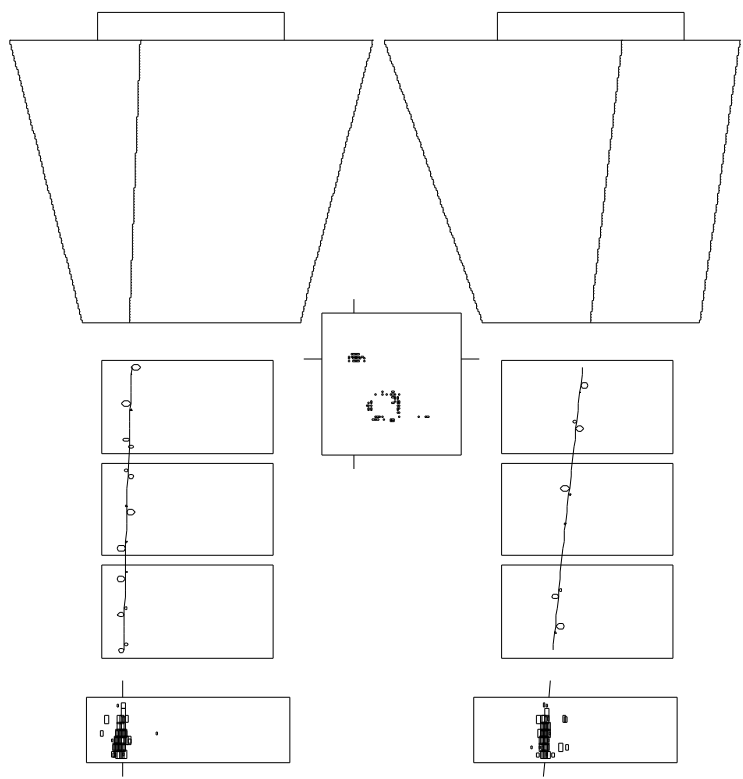}
\vspace*{-8cm}
\figcaption[f2.eps]{Display of a single 5~GV electron in the CAPRICE98 
apparatus. The instrument is shown in the view of maximum bending ({\em x})
(left) and in
the orthogonal view ({\em y}) (right).
From top to bottom is shown the RICH seen from 
the side with
a view of the signals in the pad plane (square in the
center), the tracking
stack of three drift chambers and the imaging
calorimeter at the bottom.
Note that the figure is not to scale. The calorimeter
is significantly thinner than it is shown in the figure.
The RICH shows the detected Cherenkov ring typical of
a $\beta \simeq 1$ particle
well separated from the ionization cluster of pads.
A line is drawn through all instruments that is the fitted track
 of the particle. In the drift chambers along that line there are small circles
 drawn for each wire that gave a signal. The size of the circle is
 proportional to the calculated drift time for the electrons at that wire.
The
calorimeter shows the typical signature of an electromagnetic shower
induced by the electron.
\label{elet} }
\end{figure}

Figure~\ref{elet} illustrates the calorimeter performance and shows a
schematic view of a 5~GV electron in the CAPRICE98
apparatus.
In the figure there are a left- and a right-view, symbolizing 
respectively 
the {\em x} and {\em y} views of the CAPRICE98 apparatus. At the top
is the RICH detector and a rotated 
view of the signals in the pad plane of the
multiwire proportional chamber is showed in the square frame in the
center of the figure. The ionization cluster of pads
can be seen well separated
from the Cherenkov ring typical of a $\beta \simeq 1$ particle.
The three central boxes are the drift chambers of the
 tracking system. At the bottom there is a rectangular frame
that symbolizes the
 calorimeter. The line that is drawn through all detectors represents
the fitted track
 of the particle. Along the line in the drift chambers 
there are small circles
 drawn around
each wire that gave a signal. The size of the circle is
 proportional to the calculated drift time for the electrons at that wire. The
calorimeter shows the electromagnetic shower induced by an 
electron.

\begin{figure}
\plotone{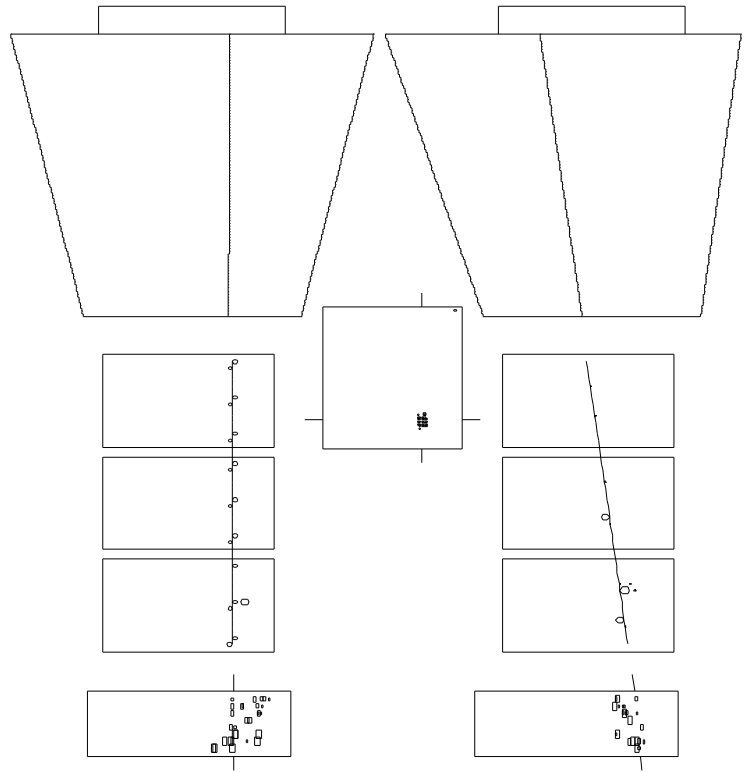}
\vspace*{-8cm}
\figcaption[f3.eps]{Display as in Fig.~\ref{elet} of a 6.4~GV antiproton 
traversing the CAPRICE98 apparatus. No Cherenkov light is detected in the RICH
pad plane where the ionization cluster can be 
seen. The antiproton interacts in the calorimeter, showing several charged 
particles emerging from the vertex of interaction.
\label{anti2}}
\end{figure}

\begin{figure}
\plotone{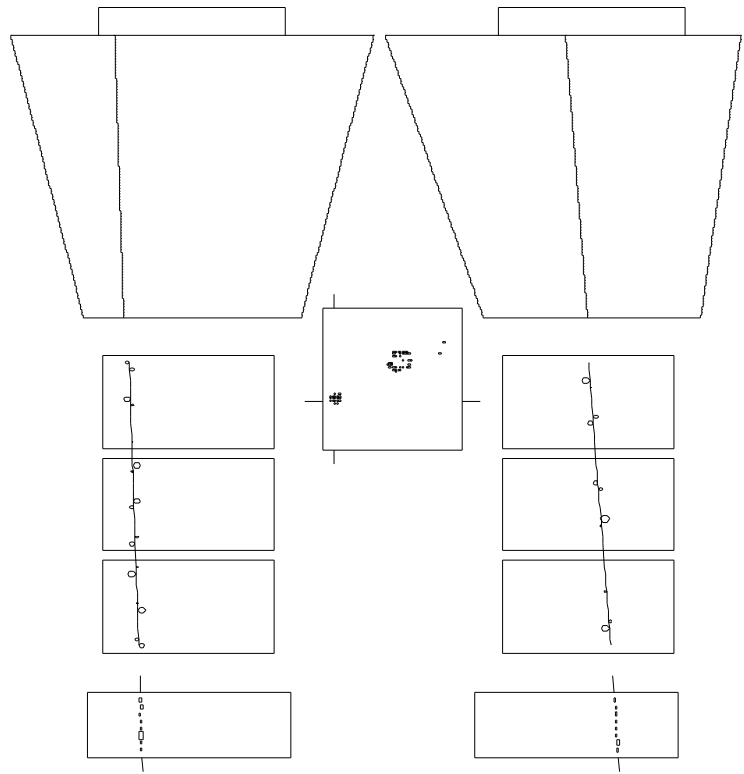}
\vspace*{-8cm}
\figcaption[f4.eps]{Display as in Fig.~\ref{elet} of a 22.7~GV antiproton 
traversing the CAPRICE98 apparatus. The calorimeter shows the typical 
pattern of a non-interacting particle.
\label{anti}}
\end{figure}

The electromagnetic shower in the calorimeter of Figure~\ref{elet} is clearly
distinguishable from the hadronic shower produced by an interacting
antiproton candidate shown in Figure~\ref{anti2} and from the 
non-interacting pattern of another antiproton
candidate shown in Figure~\ref{anti}.

Out of the 31 antiproton events surviving all antiproton 
selections 8
were found to have interacted in the calorimeter. This is
in agreement with the simulated expectation of $9.7 \pm 1.7$ antiproton
interactions in the calorimeter.

\subsubsection{RICH}
The RICH was used to measure the Cherenkov angle of the particles and
hence their velocities. Below the threshold rigidity for antiprotons to
produce Cherenkov light in the gas (about 18~GV)
the detector was
used as a threshold device to veto lighter particles, 
while above it the Cherenkov angle was
reconstructed.
Below 25~GV the fluctuations in the number of detected
photoelectrons were quite large due to 
variations in threshold rigidity caused by pressure variation
in the radiating gas and because of large Poisson fluctuations 
as the average number of photoelectrons detected 
at this rigidity was 6. Therefore
antiprotons in the rigidity range 4 to 25 GV were selected
if the event did not produce a Cherenkov signal or when  
the reconstructed
Cherenkov angle was consistent with that of an antiproton with the measured
rigidity.

\begin{figure}
\plotone{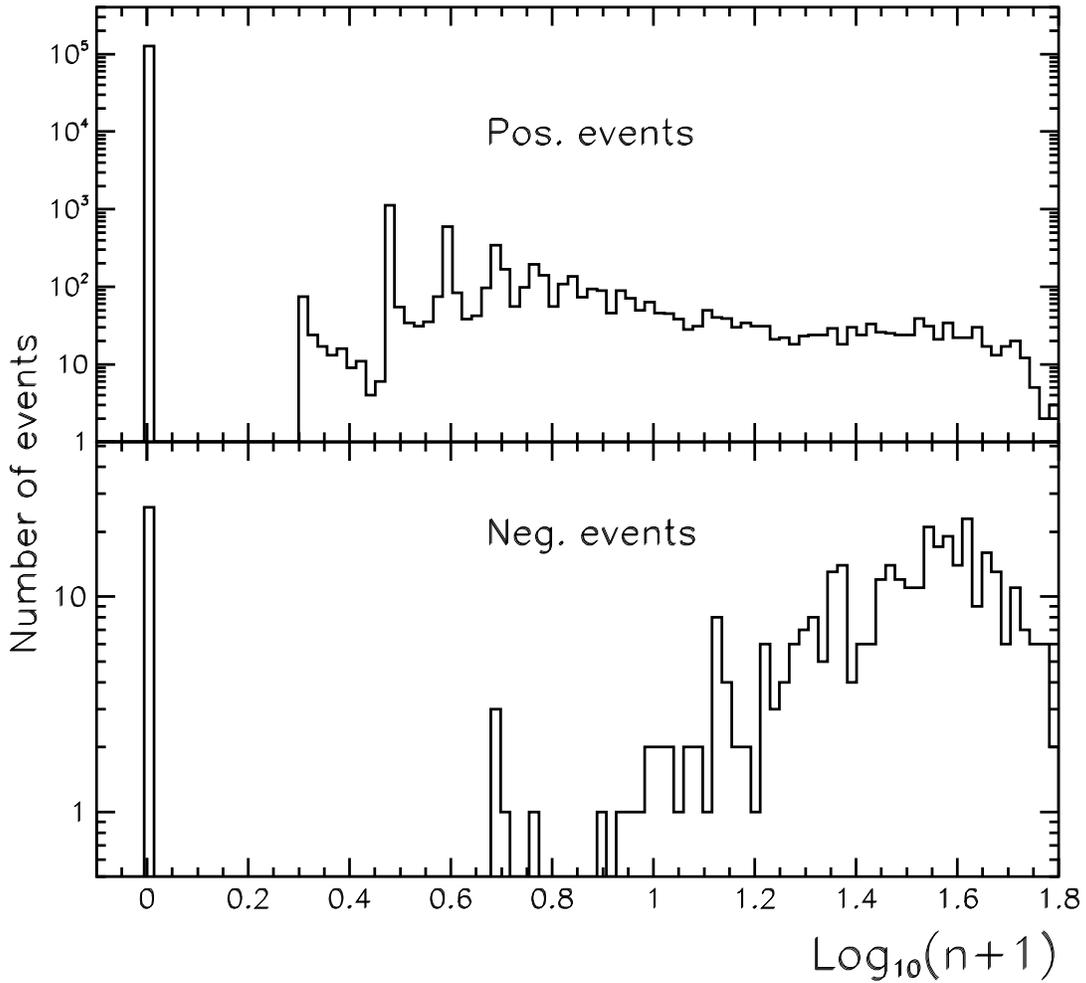}
\figcaption[f5.eps]{The distribution of logarithm with base 10 of
the number ($n$) of
pad used in the Cherenkov angle calculation
plus 1 for positive and negative particles passing the tracking, ToF and
calorimeter selection criteria in the rigidity range
from 4 to 18~GV.
\label{neff}}
\end{figure}

We show in Figure~\ref{neff} the separation between
antiprotons and lighter particles, after applying tracking, ToF
and calorimeter selection criteria,
by means of the RICH information in the
rigidity range from 4 to 18~GV. The events in Figure~\ref{neff} are plotted
as a function of the logarithm in base 10 of the number of
pads ($n$) used for the reconstruction of the Cherenkov
angle (Bergstr\"{o}m 1999\footnote{licentiat thesis 1999,
Royal Institute of Technology, is available at:
http://www.\-particle.\-kth.\-se/group\_docs/\-astro/\-research/\-references.html};
Bergstr\"{o}m et al. 2000b) plus 1.
It can be noted from the top panel that
protons have low values of $n$ (note the logarithmic scale on the
Y axis), 93\% being at zero, while on the negative side a corresponding peak
can be seen indicating the antiprotons. It is evident from the figure that
faster particles (mostly muons and pions) at higher $n$ can clearly be
separated from the antiprotons. In the analysis,
(anti)protons below 25~GV were selected by 
requiring
a value of $n$ equal to 0. In this way 26 antiprotons were identified.
These events had rigidities between 4 and 17~GV. Note that events were 
also selected if the Cherenkov angle could be constructed and 3 events were 
identified below 25 GV.

It is worth noting that when using the RICH detector as a threshold device
the tracking system gave information about
where in the pad plane the Cherenkov light should have been detected,
thereby greatly enhancing the detector immunity to noise. 
The number of noisy pads per event was, on the average, less than one out 
of 4096 channels/pads. This allowed a
stringent selection to be applied for antiprotons where no 
signals should appear in the predicted area, and yet  
maintaining a high identification efficiency. All 4096
channels were working during the entire flight.

Figure~\ref{anti2} shows the
schematic view of a 6.4~GV selected
antiproton. At this rigidity the antiproton does not produce light and, in 
fact, only the signal from the ionization due to the crossing particle
is detected in the pad plane.

\begin{figure}
\plotone{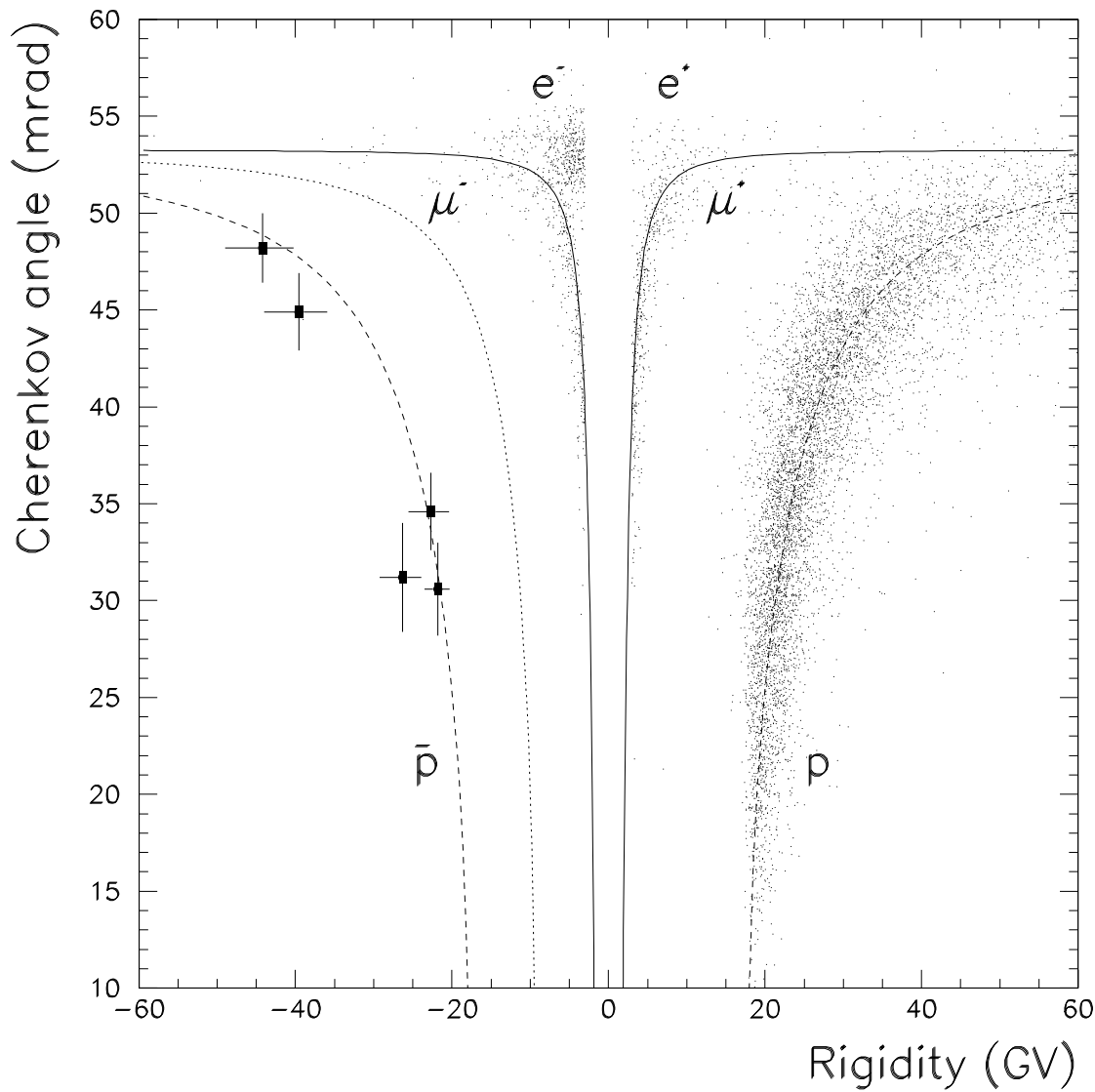}
\figcaption[f6.eps]{The measured Cherenkov
 angle for singly charged particles passing the tracking and ToF
selection criteria (8086 events) as a function of rigidity.
 The solid, dotted and dashed lines
 represent the theoretical values of the Cherenkov angle for muons,
 kaons and (anti)protons 
 respectively. To the right is a dense band of protons starting at
 approximately 18~GV and extending to higher energies and increasing Cherenkov
 angles. The main bulk of electrons and positrons were located at the low
 energies (below 10~GeV) and at maximum Cherenkov angle. On the negative side,
 the location of five antiprotons between 20 and 50~GV are indicated with
 black squares together with one standard deviation
errors on the measured rigidities and Cherenkov angles.
\label{thcvsrig} }
\end{figure}

Since the RICH detector played a crucial role in the rejection of
 background muon and pion events, stringent selection criteria were applied to
 the RICH data for the events above the antiproton threshold. The Cherenkov
 angle resolution for protons was determined using a large sample of protons
 selected using the calorimeter and scintillators. The resolution varied from
 11~mrad at threshold to about 1.3~mrad for fully relativistic protons
 \cite{ber00b}.
Figure~\ref{thcvsrig} shows the measured Cherenkov angle as a function of
rigidity. The
 events were selected from the flight data after applying tracking and ToF
selection criteria. The solid, dotted and dashed
lines indicate the calculated Cherenkov angle for  
muons, kaons and (anti)protons respectively.
Five antiprotons are clearly identified and they are
shown with black boxes and with one standard
deviation error bars for both the rigidity and
Cherenkov angle measurements.

In conclusion the conditions on the RICH detector information used for the
antiproton selection were:
\begin{enumerate}
   \item The center, extrapolated from the tracking information,
   of the Cherenkov ring was required to be
   contained in the pad plane.
   \item
   Multiple charged tracks traversing the MWPC were rejected by requiring
   that there was only one cluster of pads with a high signal, 
   typical of ionization from a charged particle in the location indicated
   by the tracking system.
   \item If the particle crossed the MWPC (46\% of the events), 
   a good agreement between the particle's impact position as determined
   by the RICH and the tracking system was required. The difference
   in x and y should be less than 3 standard deviations (rigidity
   dependent), typically $<$ 5 mm.
   \item Conditions on the Cherenkov angle for events: \\
   (i) Between 4 and 25~GV, there was no signal due to
   Cherenkov light. \\
   (ii) However, the following criteria were met 
   for all rigidities above the antiproton threshold 
   (calculated on an event by
   event base according to the measured gas pressure \citep{ber00b}):
        \begin{enumerate}
           \item A rigidity dependent condition on the number of pads
           used for the
           reconstruction of the Cherenkov
           angle was applied. The condition required  
           more than 5 pads at the
           antiproton threshold increasing to 20 above 35~GV.
           \item The reconstructed
           Cherenkov angle should not deviate by more
           than 3 standard deviations
           below and 2 standard deviations
           above from the expected Cherenkov angle for (anti)protons.
           \item To suppress the background from lighter particles, the
           reconstructed Cherenkov angle was required to be more than
           4 mrad (3 standard deviations for $\beta \simeq 1$ particles)
           away from the
           expected Cherenkov angle for pions
           (about 53~mrad above 18~GV).
        \end{enumerate}
    Hence between the antiproton threshold (about 18~GV) and 25~GV,
    antiprotons were selected
    either if the event did not produce a Cherenkov signal or if the
    reconstructed Cherenkov angle was consistent with that of an antiproton
    with the measured rigidity.
\end{enumerate}

Figure~\ref{anti} shows the
schematic view of one of the selected
antiprotons. The clean Cherenkov ring in the RICH pad plane is well separated
from the ionization cluster of pads, hence permitting the mass 
associated with the event to be reconstructed.

\subsection{Contamination}

The contamination due to e$^{-}$, $\mu^{-}$, $\pi^{-}$ and spillover protons
in the antiproton sample was studied carefully using
simulations and experimental data taken during the flight
and on the ground before the flight.

\subsubsection{Albedo particles}
Albedo particles were rejected using the time-of-flight information.
With a time-of-flight between the top and bottom scintillators of more
than 4 ns, the 0.23 ns resolution ensured a negligible background of
upgoing particles.

\subsubsection{Electron contamination}
The calorimeter performance was studied primarily with simulations.
Simulation studies showed that electron contamination in the calorimeter
selection was of ($0.6 \pm 0.2$)\% independent of rigidity in the interval
from 4 to 50~GV.
This value was cross-checked
by studying electrons selected using a condition on the total number of
strips hit in the calorimeter. It is worth noting that only the calorimeter
was able to separate electrons from muons above about 5~GV.

The electron contamination in the RICH selection was studied using a sample
of 495 e$^{-}$ in the interval 4 to 50~GV (478 between 4 and 18~GV),
selected using the
calorimeter. Of the 495 events one was selected as antiproton resulting
in an electron contamination of  ($0.20^{+0.47}_{-0.17}$)\% in the 
RICH selection for antiprotons.

To estimate
the electron background, we considered the events with negative curvature
between 4 and 50~GV. After imposing the tracking and ToF selection criteria
to this sample, 1031 negative events were left. As a worst case, we assumed
that all of these events were electrons. Applying the rejecting power of the
RICH and calorimeter to this sample resulted in an electron contamination of
less than 0.1 event over the entire range of the antiproton measurement.
Hence, the electron contamination in the antiproton sample can be assumed
negligible.

\subsubsection{Muon contamination}
The calorimeter cannot separate muons from non-interacting antiprotons.
The antiproton identification in a muon background was performed by the RICH.
The muon contamination in the RICH selection was studied using
a sample of muons collected during a ground data run prior to
the launch. The fraction of muons surviving the antiproton RICH selection was
($0.36^{+0.13}_{-0.10}$)\% between 4 and 8~GV, ($0.30^{+0.18}_{-0.12}$)\%
between 8 and 18~GV, ($0.76^{+0.60}_{-0.36}$)\% between 18 and 30~GV and
($1.7^{+1.4}_{-0.8}$)\% between 30 and 50~GV.
Defining negative muons all the events surviving the
tracking, ToF and calorimeter antiproton selection criteria,
319, 108, 24 and 11 muons were selected
from the flight data between 4 and 8~GV, 8 and 18~GV, 18 and 30~GV and
30 and 50~GV, respectively.
Multiplying these numbers 
by the surviving fractions found above
and taking into account the presence of
antiprotons in the sample,
the muon contamination in the antiproton sample
was estimated to be $1.1^{+0.4}_{-0.3}$
in the first rigidity bin, $0.28^{+0.17}_{-0.11}$ in the second,
$0.15^{+0.12}_{-0.07}$ in the third and $0.10^{+0.15}_{-0.08}$ in the forth.
This contamination was
later subtracted from the antiproton signal and is shown in parenthesis in
Table~\ref{prpb}.

\subsubsection{Meson contamination}
Pions started to produce light in the gas-RICH above about 3~GV. Since the
RICH is a $\beta$-detector, the pion contamination was studied by 
scaling the muon sample. The contamination was similar to 
that of muons except in the first bin where it was
($0.6 \pm 0.1$)\%. However, pions were a small component compared to
muons in the flight data. Between 3 and 4~GV pions could be clearly
identified by the gas-RICH and it was found that pions were
($12 \pm 2$)\% of the
muons. Assuming that a similar value is valid for the bin 4 to 8~GV, the
resulting pion plus muon contamination in the antiproton sample in this bin
was $1.17^{+0.35}_{-0.27}$.

Theoretical calculations \citep{ste81} of the kaon spectrum in the
atmosphere indicate that the kaon to antiproton ratio is about 2\%.
Hence, they
are a negligible contamination of the antiproton sample. Furthermore, above
about 9~GV kaons are suppressed by the RICH selection. It is worth
pointing out that, as can be seen in Figure~\ref{thcvsrig}
(dotted line) , no negative kaons were  
identified between 9 and 50~GV.

\subsubsection{Spillover proton contamination}
\label{s:spil}
Spillover protons can represent a non negligible contamination
in the antiproton sample above 20 GV.

To obtain the shape of the spillover proton distribution a Monte Carlo
approach was
used. We started with an input power law spectrum in rigidity that was
transformed in deflection and smeared with values randomly picked from the
resolution function. The spectral index for the power law was obtained from
the proton spectrum measured by CAPRICE98 above 20~GV and was found to be:
$-2.74 \pm 0.02$.

The resolution function was obtained from a measurement of straight 
tracks taken at ground prior to the
flight without the magnetic field. These were 
analyzed as if
they were high rigidity events with magnet on. In this case the resolution
function is simply the deflection distribution. To construct this distribution
events that suffered multiple Coulomb scattering had to be eliminated since
they could enlarge the distribution
\citep[see][]{men00}.
In fact, the multiple scattering affected
events essentially at low energy while spillover events were due to high
energy protons. This was done by requiring a Cherenkov light signal 
along with high value
of the measured Cherenkov angle in the RICH so as to select close to fully
relativistic events (mostly muons and electrons). Then electrons were
eliminated using the calorimeter.
The large majority of the
selected events were particles (muons) with rigidity greater than 5 GV.

\begin{figure}
\plotone{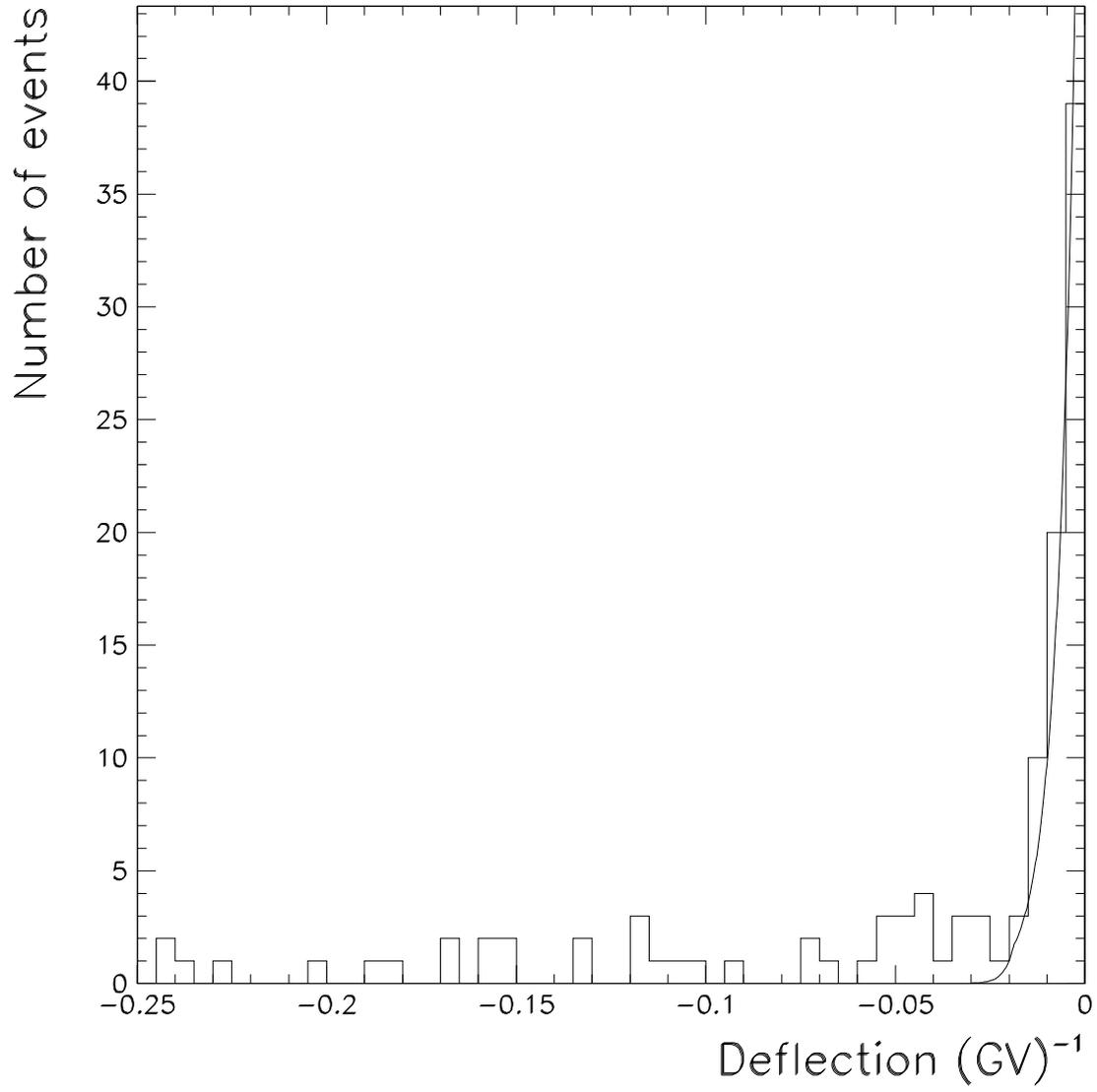}
\figcaption[f7.eps]{Deflection distribution from flight data selected
with all antiproton conditions except requirements on the Cherenkov angle.
The solid line is a fit of the proton spillover contribution.
\label{spil}}
\end{figure}

The simulated proton spillover
distribution was normalized with the experimental deflection distribution
between -0.02 and 0~(GV)$^{-1}$ and used to
obtain an estimation of the experimental
proton spillover distribution. Figure~\ref{spil} shows the experimental
distribution with the estimated proton spillover contribution (solid line).
The distribution includes spillover protons, muons and antiprotons. 
The events for the deflection distribution
were selected from the whole flight data set with the complete
antiproton selection (tracking, ToF, RICH and calorimeter conditions) 
except for the condition on the reconstructed Cherenkov angle.
At the smallest deflections (high rigidities) the dominant component was the
spillover protons.

Finally, the
contamination of spillover protons was
determined from this calculation by integrating 
the experimental spillover distribution over the proper deflection 
range. Between 0 and 30~GV the estimated
spillover contamination was less than 0.002 events and between 0 and 50~GV it
was about 0.39. However, this was not the real spillover contamination in the
 antiproton sample since antiprotons above the threshold of the RICH detector
 were selected with the additional
condition on the measured Cherenkov angle.
For the rigidity range from the threshold of the RICH detector up to
 50~GV the spillover proton contamination
 was obtained by calculating the probability that the 0.39 spillover protons
 previously estimated were selected with the Cherenkov angle condition.
This probability was obtained using negative events with
 deflection smaller than 0.01~GV$^{-1}$ (i.e. essentially spillover protons).
Using this process, we derived that the
contamination of spillover protons was 0.04 events below 50~GV.

\subsection{Efficiency}

The antiproton selection efficiencies were studied using a large sample
of experimental protons from flight data set.
It was assumed that protons and antiprotons
had the same efficiencies in the
RICH, scintillators, and the tracking system.
However, in the calorimeter
the efficiencies were not assumed to be the same because of the difference in
the interaction cross sections. Hence, the calorimeter efficiency
for protons and antiprotons was studied
using simulations.

The tracking efficiency was studied with two independent methods
(see \citet{ber00c}).
Both were based on track reconstruction
codes that were independent of the drift chamber tracking system and used 
information from the
other available detectors. The first 
method used the 
position of the
ionization cluster in the MWPC of the RICH detector, information from 
the two
ToF scintillation detectors (only in the x direction) and from
the electromagnetic calorimeter to reconstruct the track of the particle
traversing the detector system. This combination 
had an estimated MDR of 4.5 GV.
The second method used the RICH to determine the rigidity from
the velocity derived from the Cherenkov angle measurement with the
help of an extrapolated straight track from the calorimeter.

\begin{figure}
\plotone{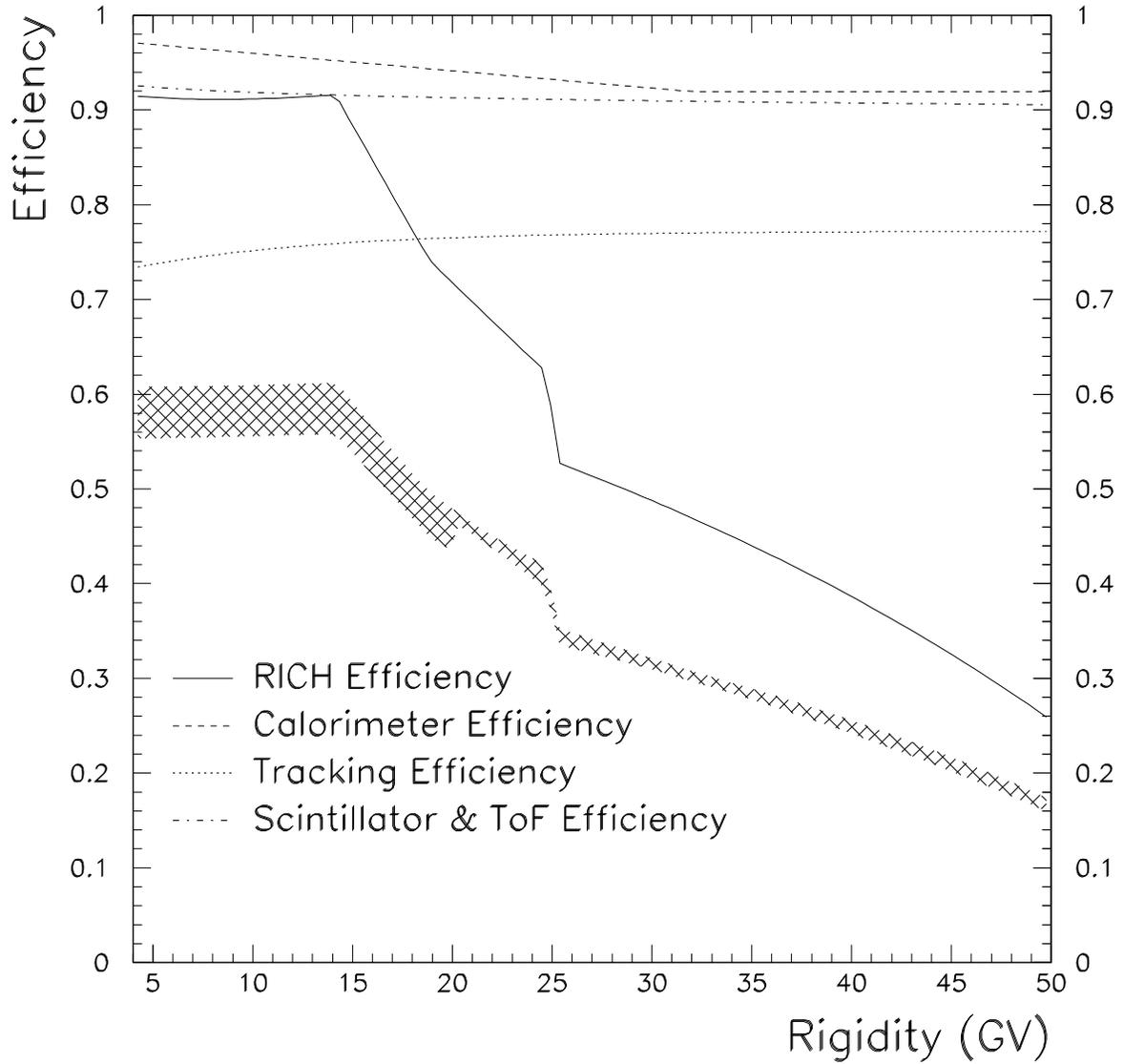}
\figcaption[f8.eps]{Efficiencies for the different detectors  
of the magnet spectrometer
for detecting antiprotons.
The hatched area indicates the one standard deviation confidence 
interval of the combined efficiency.
\label{eff}}
\end{figure}

The two methods were tested with muons from ground data and similar results
were obtained between 0.2 and 10~GV. With flight data, the first method
sampled the tracking efficiency of protons below 10~GV while the second above
18~GV (because of the gas-RICH threshold). In this case the efficiencies
differed by $\simeq 6\%$. From previous experience with a similar tracking
system~\citep{boe99} the proton tracking efficiency was expected to reach
a plateau above 2~GV.
This seems in disagreement with what found here. However,
the first method could be biased by contamination of secondary low energy
protons that would not affect the second method
(for more details see \citet{ber00c}). Thus, the
efficiency of the tracking selection was obtained by the second method
for the rigidity range from 4 to 50~GV
(dotted line in Figure~\ref{eff}). The difference found
 between the proton tracking efficiency with the two methods was
 considered a systematic uncertainty and a 6\% systematic
 uncertainty was included in the flux calculation for rigidities between
 4 and 20~GV.

Secondary particles backscattering from the calorimeter could result in an
inefficiency for the tracking selection. We studied this effect selecting 
from the flight data
events that interacted in the calorimeter producing large signals 
(above 1.8~mip) in the bottom scintillator. For these events we applied the 
tracking selection. Accounting for the relative abundance of these events,
we found an overall decrease in the tracking efficiency of about 0.4\%.

A Monte Carlo simulation based on the CERN GEANT/FLUKA-3.21 code \citep{bru94}
was used to study the calorimeter selection efficiencies.
Results from similar simulations were found to be in good agreement
with test beam data and experimental
results from previous balloon flights (Bocciolini et al. 1993;
Boezio 1998\footnote{Ph.D. thesis 1998,
Royal Institute of Technology, is available at:
http://www.\-particle.\-kth.\-se/group\_docs/astro/research/references.html}).
It was found that the efficiency for the selection of
 antiprotons and protons using simulated data were in agreement, within the
 statistical uncertainties. The calorimeter efficiency for protons
 selected from experimental data was in reasonable agreement 
 with previous simulations. 
 Differences (about 2\%) found 
 were probably due to the less
 accurate simulations of hadronic showers compared to
 electromagnetic showers.
 Since simulations indicated  
 that protons' and antiprotons' calorimeter
 selections had the same efficiency,
the efficiency of the experimental proton calorimeter
 selection was also used for the antiprotons (dashed line in Figure~\ref{eff}).

The RICH efficiency was rigidity dependent and is shown as solid line
in Figure~\ref{eff}. As expected the efficiency was constant below the
gas-RICH threshold but it started decreasing 
at 14~GV due to above threshold protons that spilled down to
lower rigidities because of the finite resolution of the tracking system.
Above 25~GV the antiprotons were selected only with conditions on the
Cherenkov angle . The
decrease of the RICH efficiency above 30~GV was caused by
the requirement that the Cherenkov angle should be more than 4~mrad
away from the expected Cherenkov angle for pions.

Since the detector efficiencies varied with rigidity,
mean efficiencies had to be
calculated. This was done by weighting the efficiencies
in each bin with the proton spectrum measured in this experiment (with
a larger number of bins and no RICH selection) and with an
antiproton spectrum given by interstellar secondary calculation
by L. Bergstr\"{o}m \& P.
Ullio (1999, private communication) which included the effects if the
geomagnetic cut-off.

\subsection{Geometrical factor}

The acceptance of the instrument allowed for particles with a range 
         of zenith angles to be measured. The maximum angle was 14 degrees 
         and the mean of the distribution was at 8 degrees.

The
geometrical factor was obtained with Monte Carlo techniques \citep{sul71}
and the same track-fitting algorithm used in this
analysis to trace the particles through the spectrometer. The
 geometrical factor ($G$) was found to be constant
at $155.0 \pm 1.1$~cm$^2$~sr
in the rigidity range from 4 to 50~GV.

\subsection{Payload and atmospheric corrections}

\subsubsection{Payload corrections}
To determine the number of particles at the top of the payload, 
the losses and the
 production of particles due to interaction in the material of the payload had
 to be considered. To reach the tracking system of the spectrometer, the
 particles had to go through first the aluminum shell of the payload, 
 the RICH detector and then 
 the top scintillator of the time-of-flight system. It
 was assumed that all particles that interacted above the tracking system
 were rejected by the selection criteria. The probability of an interaction in
 the material of the drift chambers, that would not be rejected by the
 tracking system conditions, was considered as being negligible.
The data were corrected for these losses with multiplicative factors, using
   the expression for the interaction mean free path for the different
   materials in the detectors given by \citet{ste97}.
This gave correction factors of 1.132, 1.118, 1.109 and 1.104 for antiprotons
and of 1.081, 1.082, 1.084 and 1.086 for protons in the four rigidity
intervals.
The corrected number of antiprotons and
   protons at the top of the payload are given in Tables~\ref{prpb}
and \ref{t:ratio}.

\subsubsection{Atmospheric corrections}
For the production of secondary antiprotons and protons in the atmosphere, 
we used the calculation by \citet{pap96} for the protons and 
the calculation by \citet{ste97} for the antiprotons.
To determine the secondary spectra we compared
our measured proton spectrum propagated to the top of the atmosphere with the
spectrum used for the secondary calculations at solar minimum
\citep{pap96,ste97} from which a normalization factor was derived. The
resulting secondary fluxes were 
normalized with the geometrical factor
   and live time of the experiment, and subtracted from the corrected numbers
   using a mean residual atmosphere of 5.5 g/cm$^{2}$. The number of
atmospheric antiprotons and protons are given in Tables~\ref{prpb}
and \ref{t:ratio}.

The correction for losses in the atmosphere was carried out 
using a method
analogous to the instrument correction. This gave correction factors of
1.1, 1.089, 1.082 and 1.079 for antiprotons and of
1.061, 1.063, 1.064 and 1.065 for protons
in the four rigidity intervals.

\subsection{Geomagnetic transmission correction}
To be able to get the fluxes at the top of the atmosphere,  
the transmission
of the particles through the earth's magnetic field had to be taken into
account. During the flight the position of the payload changed between
34.3$^{\circ}$ and 35.5$^{\circ}$ North Latitude and
between 104.1$^{\circ}$ and
110.1$^{\circ}$ West
Longitude. This correspond to an average value of the vertical
cut-off rigidity of about 4.3~GV \citep{she83}.
However, this cut-off is not a sharp value
below which all particles are deflected and cannot reach
the apparatus and above
which all particles arrive. In fact, around the geomagnetic cut-off the
particles are partially transmitted through the earth magnetic field.
Furthermore, the penumbral bands define forbidden bands of rigidity, which vary
with arrival direction, time and geographic location. In this analysis
all these effects are
represented by a transmission function which was derived by the experimental
data.

We found that the CAPRICE94 \citep{boe99} and CAPRICE98 proton spectra
above about 10~GeV are
nearly identical in shape and the absolute fluxes differ by $\simeq
7\%$ in good agreement considering both the
statistics and systematic errors, which in \citet{boe99} were estimated to be
of the order of 10\%. Moreover, the solar modulation during the two balloon
flights was also very similar. The values from the
neutron monitor counter CLIMAX\footnote{National Science
Foundation grant ATM-9912341,
http://ulysses.uchicago.edu\-/NeutronMonitor\-/Misc\-/neutron2.html} 
\citep{sim96} 
were 415600 counts/hour and
417000 counts/hour at the time of the CAPRICE94 and CAPRICE98 flights,
respectively.
However, the CAPRICE94 experiment took place in North
Canada at an average geomagnetic cut-off of about 0.5~GV. Hence, above 1~GV
the effects of the geomagnetic field on the CAPRICE94 proton spectrum were
negligible. Consequently, the transmission function was defined
as the ratio between the experimental
CAPRICE98 and CAPRICE94 proton fluxes.

The correction factors ($TF$)
for the geomagnetic effect were derived weighting the
transmission function with the proton and antiproton spectra
as done for the efficiencies. The resulting
transmission values differ from 1 only in first bin (4 to 8~GV) where they are
$0.84 \pm 0.06$ and $0.81 \pm 0.07$ for antiproton and protons, respectively.

\section{Results}

\subsection{Antiproton flux at the Top of the Atmosphere}

Given the number of events ($N^{TOA}_{\ap}$) selected with the antiproton
criteria and corrected for selection efficiencies, losses in the payload and
in the atmosphere and atmospheric secondaries,
we obtained the antiproton fluxes at the top of the
atmosphere from the relation,
\[
   {\mbox {\rm Flux(E)}} \: = \: \frac{1}{T_{live} \times G \times \Delta E
   \times TF}
   \times N^{TOA}_{\ap}(E) ,
\]
where $\Delta E$ is the energy bin corrected for ionization
losses to the top of the atmosphere and
$E$ the kinetic energy.
   The resulting antiproton flux is given in Table~\ref{fluxes}. The total
   errors include both statistical and systematic errors.
The mean energies of the bins are given according to \citet{laf95}.

Figure~\ref{fluxpb} shows the antiproton flux measured by this
 experiment together with other experimental data
 \citep{buf81,mit96,boe97,bas99,ori00}.
 The two solid lines show the upper and lower limit of a calculated flux of
 interstellar secondary antiprotons \citep{sim98} using a recently
measured proton and
 helium spectra \citep{men97} and a reacceleration model which allows
energy-changing
 processes caused by the non-annihilation process and by elastic scattering.
 The dashed line shows the interstellar secondary antiproton flux
calculated by L. Bergstr\"{o}m \&  P.
Ullio (1999, private communication). This calculation assumed a
diffusion model of propagation with
 an isotropic diffusion coefficient and no reacceleration. It used the
 interstellar proton spectrum measured by the CAPRICE94 experiment
 \citep{boe99}. The dotted line shows the primary
 antiproton flux by \citet{ull99c}, which included a Minimal
 Supersymmetric Standard Model with a contribution from an assumed
 Higgsino-like neutralino, with a mass of 964~GeV.
The theoretical fluxes, but not the experimental values of the other
experiments, were corrected for the solar
   modulation conditions corresponding to the CAPRICE98 flight
using a spherically symmetric model \citep{gle68} with a solar
 modulation parameter of $\Phi = 600$~MV.

\subsection{Antiproton to Proton ratio}

To obtain the antiproton to proton ratio at the top of the atmosphere,  
we corrected the number of selected antiprotons and protons
for the production and loss of particles in the
residual atmosphere above the apparatus as well as in the instrument itself.
However, for obtaining antiproton to proton ratio, 
the selection efficiencies, which were considered to be the same for
antiprotons and protons, were excluded from the calculation
in order to reduce the errors. The resultant ratios are presented in
Table~\ref{t:ratio}.

\begin{figure}
\plotone{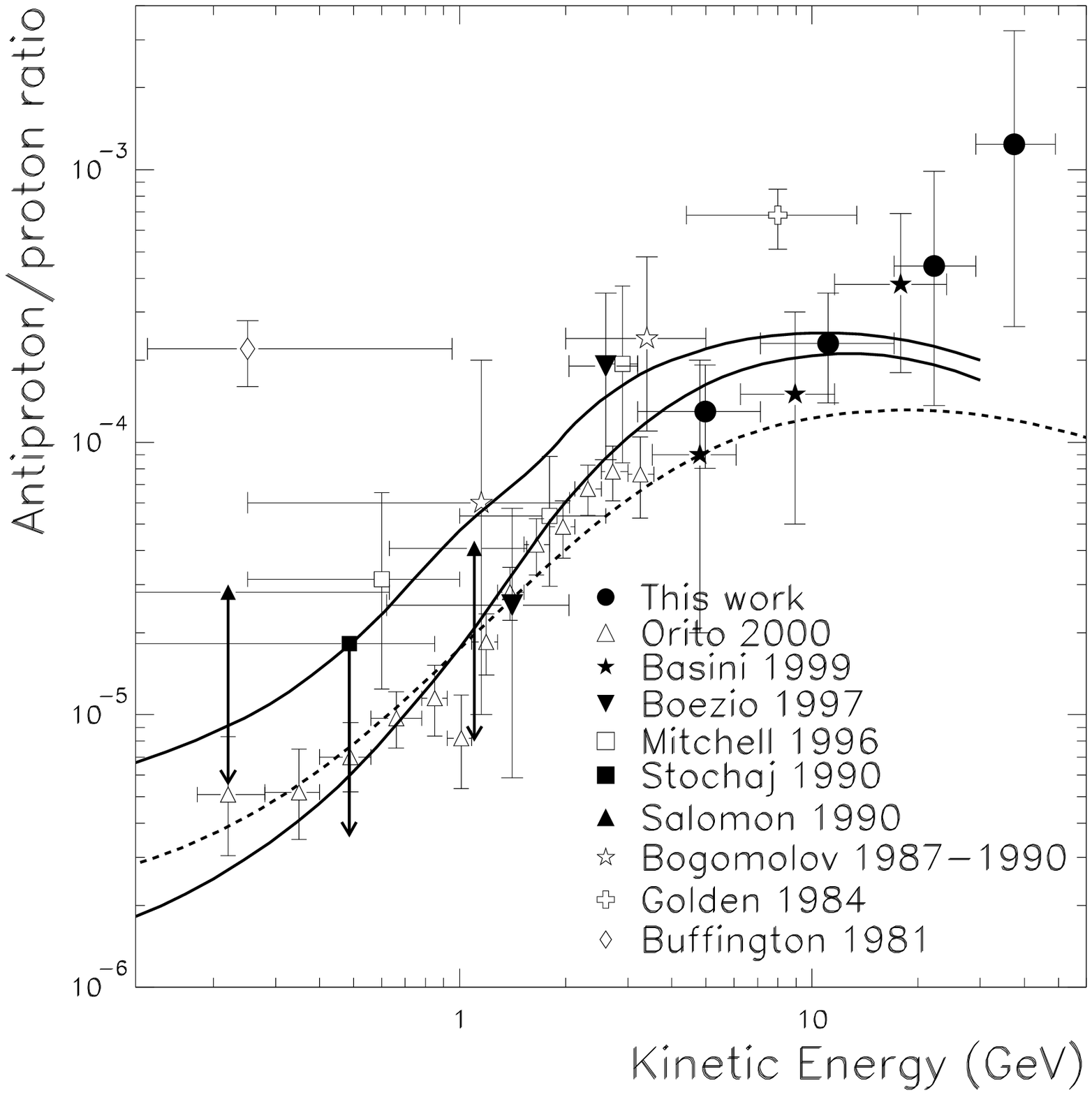}
\figcaption[f10.eps]{The \ap/p ratio at the top of
the atmosphere obtained in
this work compared with previous measurements
\protect\citep{buf81,gol84,bog87,bog90,sal90,sto90,mit96,boe97,bas99,ori00}.
The lines are the calculations of interstellar antiprotons
assuming a pure secondary production
 during the propagation of cosmic rays in the Galaxy by \protect\citet{sim98}
(solid lines, upper and lower limits of the calculation)
and by L. Bergstr\"{o}m and P.
Ullio 1999 (private communication) (dashed line).
\label{ratio} }
\end{figure}

Figure~\ref{ratio} shows the antiproton to proton ratio measured by
 CAPRICE98  along with other experimental data
\citep{buf81,gol84,bog87,bog90,sal90,sto90,mit96,boe97,bas99,ori00} and with
 theoretical calculations. The two solid lines show the upper and lower limit
 of a calculated flux of interstellar antiprotons by \citet{sim98}
assuming a pure secondary production
 during the propagation of cosmic rays in the Galaxy .
The dashed line
shows a similar calculation by L. Bergstr\"{o}m and P.
Ullio (1999, private communication). It is worth noting that 
\citet{sim98} used the primary spectra measured by \citet{men97} and that
L. Bergstr\"{o}m and P.
Ullio (1999, private communication) used the 
interstellar proton spectrum measured by \citet{boe99}.

The antiproton
to proton ratio values presented here are in
perfect agreement with
the preliminary CAPRICE98 results published by \citet{ber00a}. A
small variation, of about a third of standard deviation, is found
only for the highest energy bin essentially due to a better
understanding of the proton spillover contamination.

\section{Systematic uncertainties}
\label{systematics}

Systematic errors originating from the determination of the detector
efficiencies have already been discussed and they have been
included in the data in the figures and in the tables.

Another possible systematic error is related to the efficiency of the trigger
system. The trigger efficiency was studied during the pre-flight preparations
with a system measuring the coincidence of two scintillators placed above and
below the top and bottom time-of-flight scintillators. The efficiency was
found to be close to 100\% with an uncertainty of about 2\%. The performance
of the trigger system during the flight was also studied comparing the
experimental spatial distribution of triggers with the distribution given by
the same simulations as were used for the geometrical factor calculation, and
an excellent agreement was found. Hence, the trigger efficiency could be
assumed to be 100\%, with a possible systematic uncertainty of less than 2\%.

The method for calculating the geometrical factor used in this work was
compared with two other techniques in the CAPRICE94 
analysis~\citep{boe99}, 
and it was found to be in agreement within 2\%, above 0.5~GV. Considering the
similar geometrical configuration of CAPRICE98, it can be concluded that the
uncertainty on values of the geometrical factor was about 2\%.

Systematic errors due to the uncertainty on the secondary production of
antiprotons in the atmosphere were estimated comparing the results from the
calculation by \cite{ste97}, used in this work, with the independent
calculation by \cite{pfe96}. It was found that the two calculations
differ by $\simeq 22\%$ at 5 GeV decreasing to $\simeq 2\%$ above 8~GeV, with
the calculation by \cite{pfe96} being the lower. This introduces an
estimated
uncertainty in the antiproton fluxes extrapolated to the top of the
atmosphere, that is $\simeq 6\%$ at 5~GeV decreasing to less than 1\% above
8~GeV.

The atmospheric secondaries were also affected by the uncertainty in the
residual atmosphere above the gondola. This was measured to be 5.5~g/cm$^{2}$
by a pressure sensor owned and calibrated by the CAPRICE collaboration. The
pressure was also measured by a detector owned and calibrated by
the National Scientific Balloon Facility (NSBF).
The NSBF pressure data were about 15\% lower at float than the ones
measured by our sensor. This results in an uncertainty on the antiproton
fluxes of about 7\% between 3 and 20~GeV decreasing to about 2\% above
30~GeV.

The numbers of particles measured at the spectrometer were corrected for
losses in the spectrometer and the atmosphere. Assuming a 10\% uncertainty on
the cross sections used in these calculations results in a systematic error
on the antiproton fluxes of $\simeq 2\%$. An additional uncertainty of 1\%
should be considered due to the uncertainty on the atmospheric depths and the
consequent effect on the losses in the atmosphere.

Assuming that the systematic errors were uncorrelated and, hence, could be
quadratically summed, the resulting measurement of the antiproton flux
includes systematic uncertainties that were energy dependent decreasing from
$\simeq 10\%$ at 5~GeV to $\simeq 8\%$ above 8~GeV and to $\simeq 4\%$ above
30~GeV.

Systematic uncertainties due to the tracking system, caused by, for example, 
an offset in the deflection measurements, were analyzed using the
RICH detector.
The RICH high Lorentz threshold for protons permitted to study
several features of the tracking system up to a rigidity of about 100~GV
(unattainable by previous cosmic rays experiments).

\begin{figure}
\plotone{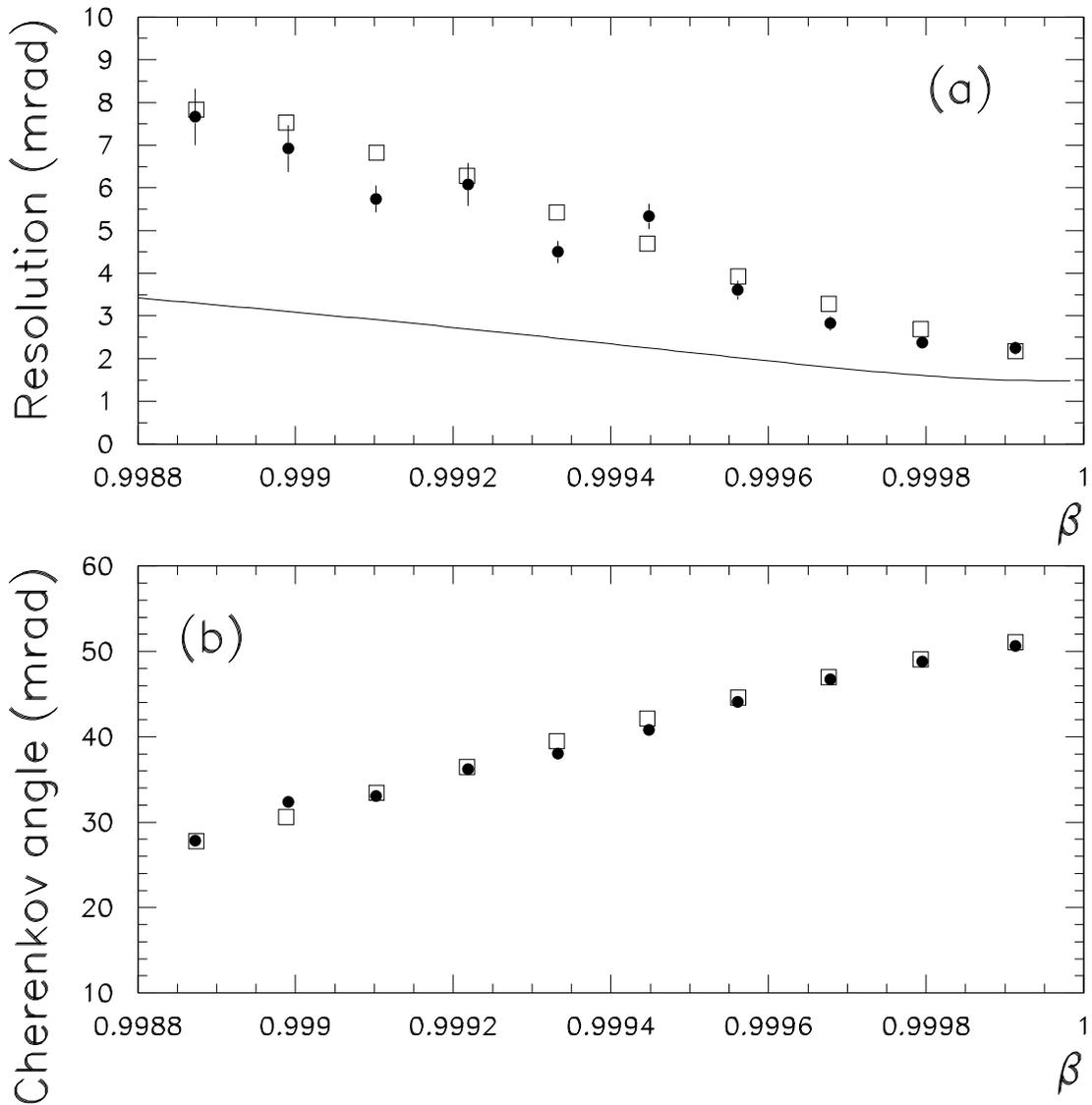}
\figcaption[f11.eps]{(a) Cherenkov
 angle resolution as a function of velocity for experimental ($\bullet$)
 and simulated ($\Box$) protons. The solid line is the experimental resolution
 for muons. (b) Mean Cherenkov angle as a function of velocity for 
experimental ($\bullet$) and simulated ($\Box$) protons.
\label{richtrack}}
\end{figure}

Figure~\ref{richtrack}(a) shows the Cherenkov angle resolution for protons
obtained from flight data ($\bullet$) as a function of $\beta$,
derived from the rigidity measured by the tracking system assuming the proton
mass. The solid line indicates the measured resolution for muons
\citep{ber00b}. The
difference between the two resolutions was due to the finite resolution of
the tracking system. In fact, this introduced an additional spread in the
Cherenkov
angle distribution, when it was binned according to the $\beta$ measured by
the tracking system. The effect of this was more important for protons than
for muons, since the rigidity (deflection) of the protons was nearly 10 times
greater (smaller) than that of the muons, at the same $\beta$.

The effect of
the tracking resolution of the binned Cherenkov angle distribution was
obtained by simulating a large number of protons according to a power law
spectrum in rigidity with spectral index of -2.74.
The
corresponding Cherenkov angle was then calculated
and smeared with a gaussian distribution with a 
standard deviation given by the measured resolution for muons 
(Fig.~\ref{richtrack}(a), 
solid line). Then the rigidity of each simulated proton was transformed
to a corresponding
deflection and smeared with values randomly picked from the tracking
resolution function. The resulting deflection was
then used to derive the velocity, which was used for the binning of the
Cherenkov angles of these simulated events, similarly to the experimental
case. The resulting Cherenkov angle resolution is shown as $\Box$ in
Figure~\ref{richtrack} (a). A good agreement was found between the
measured and simulated resolutions. This was a strong indication that the
tracking resolution function used in this work describes the tracking 
uncertainties with high precision and, 
consequently, could be used
for determining the spillover proton contamination in the antiproton sample.

Figure~\ref{richtrack}(b) shows the measured mean Cherenkov angle ($\bullet$)
as a function of $\beta$ from the tracking system along with the simulated
($\Box$) one. The same simulation was used as for the Cherenkov angle
resolution. Also in this case, an excellent agreement was found. Furthermore,
the comparison between the two sets of mean Cherenkov angle permitted to
limit the possibility of an offset in the measured deflection, due to effects
such as: wrong alignment of the drift chambers, positioning of the
center of the magnet, mapping of the magnetic field, etc. 
An offset was
introduced in the simulated deflection, and it was varied over a wide range
of possible values. It was found that, if an offset existed, at a 95\%
confidence level it was not larger than 0.001~(GV)$^{-1}$, which was
significantly smaller than 0.003~(GV)$^{-1}$, corresponding to the MDR of the
experiment (330~GV).

It is worth pointing out that the simulation was tested also with different
spectral index such as: -2.6 and -2.8, and no significant variation from the
case presented here was found.

\section{Conclusion}

\begin{figure}
\plotone{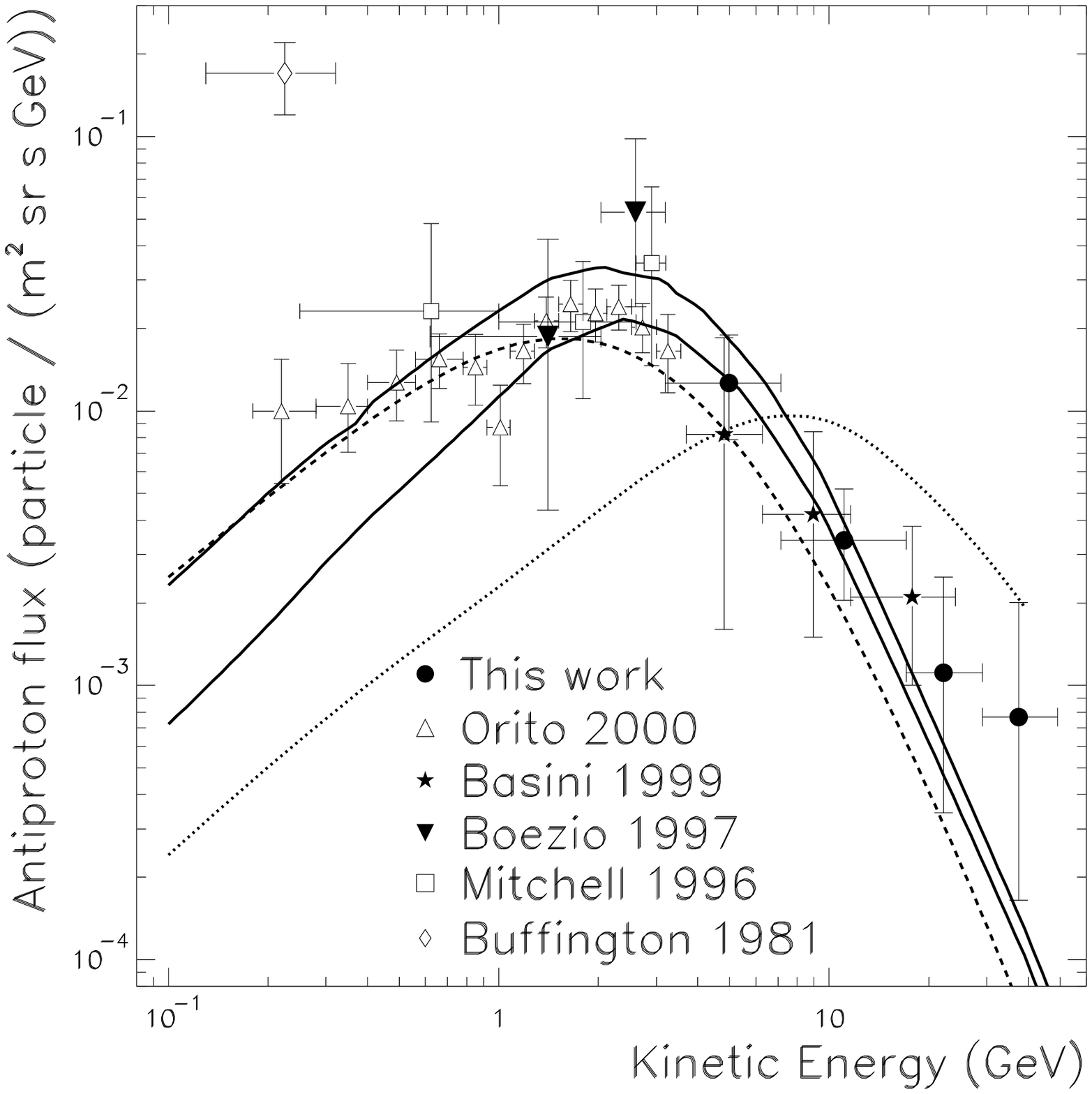}
\figcaption[f9.eps]{The antiproton flux at the top of the atmosphere
obtained in this work and compared to other experiments that have
published results on the antiproton flux
\protect\citep{buf81,mit96,boe97,bas99,ori00}.
The two solid lines shows the upper and lower limit of a calculated flux of
 interstellar secondary antiprotons by \protect\citet{sim98}.
The dashed line shows the interstellar secondary antiproton flux
calculated by L. Bergstr\"{o}m \& P.
Ullio (1999, private communication). The dotted line shows the primary
 antiproton flux given by annihilation of neutralino from MSSM with a
mass of 964~GeV \protect\citep{ull99c}.
\label{fluxpb} }
\end{figure}

The antiproton flux and the antiproton to proton ratio have been
determined in the energy region from
3 to 49 GeV by the CAPRICE98 experiment. This is the first time that the
antiproton flux has been measured up to such high energies and over such a
wide range in energy. Between 3 and 20~GeV our antiproton fluxes
are consistent with the
measurement by the MASS91 experiment \citep{bas99}. Both of these
results, within the
experimental errors, are also in agreement with the theoretical predictions by
\citet{sim98} and L. Bergstr\"{o}m and P.
Ullio 1999 (private communication) which assume a
purely secondary origin of the cosmic-ray antiprotons.
However, a primary component cannot be excluded and
the shape of the measured antiproton flux could indicate a presence of
primary antiprotons. In fact, in the CAPRICE98 analysis
we observed two antiproton events, with the
highest energy antiproton measured at a kinetic energy of 43~GeV,
between 29 and 49~GeV, while the expected number from a pure secondary 
origin is 
only 0.2 to 0.4 events, including muon contamination from 
the atmosphere; the lower and upper values correspond to 
the two extreme secondary curves of Figure~\ref{fluxpb}.
It is essential to improve the statistics on
antiproton measurements in this high-energy region
In fact, the energy region studied here  
permits to search for specific signatures of
neutralino-induced antiproton fluxes which are not attainable in lower energy
regions. Furthermore, 
in this energy range nearly all calculations of
interstellar secondary
antiprotons in the literature are consistent with each other \citep{ull99c}.
These data are also substantially
free of uncertainties due to solar modulation effects.

As a continuation of its ballooning activity, the WIZARD collaboration
   has developed a cosmic-ray experiment, named PAMELA \citep{adr99}, 
   that will be launched in quasi-polar orbit on board of a Russian satellite 
   at the beginning of 2003. PAMELA is based on a magnetic spectrometer with 
   an MDR exceeding 800~GV and will determine the antiproton spectrum from 
   80~MeV, with a few orders of magnitude better statistics at energies above 
   5~GeV than the existing ones, free of the atmospheric background.
   Another cosmic-ray experiment, AMS \citep{ahl94}, dedicated to the
   search for antinuclei, will be installed sometime later on the 
   International Space Station. As it will have a similar MDR and will be 
   based on the same set of detectors as PAMELA, it also will give further 
   results on the study of high energy antiprotons. Its much larger acceptance 
   also will allow improvement to the statistics in the same high energy range 
   that will be explored by PAMELA.

\acknowledgments
This work was supported by NASA Grant NAGW-110, The Istituto Nazionale
di Fisica Nucleare, Italy, the Agenzia Spaziale Italiana, DARA and
DFG in Germany, the Swedish National Space Board and
the Knut and Alice Wallenberg foundation.
The Swedish-French group thanks the EC SCIENCE programme for support.
We wish to thank the National
Scientific Balloon Facility and the NSBF launch crew that served
in Fort Sumner. We would also like to acknowledge the essential support
given by the Gas Work Group (EST/SM/SF) and the Thin Films
\& Section (EP/TA1/TF) at CERN, the LEPSI and
CRN-Strasbourg and the technical staff of NMSU and of INFN.

\clearpage

\begin{deluxetable}{ccccc}
\tablewidth{0pc}
\tablecaption{Summary of antiproton results
\label{prpb}}
\tablehead{
Rigidity at the & Observed number\tablenotemark{a} & Extrapolated number  &
Atmospheric  & Extrapolated number \\
spectrometer & of events   &
at top of  & secondaries & of
primary  \\
(GV) & at spectrometer & payload & & events at TOA}
\startdata
4.0-8.0 & 15(1.17) & 25.9$^{+9.6}_{-7.3}$  & 6.73  &
21.1$^{+10.5}_{-8.0}$ \\
8.0-18.0 & 11(0.28) & 20.1$^{+8.4}_{-6.1}$  & 4.43  &
17.0$^{+9.2}_{-6.7}$ \\
18.0-30.0 & 3(0.15) &  7.4$^{+7.7}_{-4.3}$  & 1.16  &
6.8$^{+8.3}_{-4.7}$ \\
30.0-50.0 & 2(0.14) &  7.8$^{+11.6}_{-5.7}$ & 0.547 &
7.8$^{+12.5}_{-6.1}$
\enddata
\tablenotetext{a}{The numbers shown in parenthesis in column 2 are the
estimated background events due to muons, pions and, in the highest energy 
bin, also spillover protons.}
\end{deluxetable}

\clearpage

\begin{deluxetable}{ccc}
\tablewidth{0pc}
\tablecaption{Antiproton fluxes at the top of the atmosphere (TOA)
\label{fluxes}}
\tablehead{
Kinetic Energy at  & Mean Kinetic Energy & Antiproton flux at
TOA \\
TOA (GeV) & at TOA (GeV) & (m$^{2}$ sr s GeV)$^{-1}$ }
\startdata
3.19-7.14   & 4.97  & $(12.6^{+6.3}_{-4.8})   \times 10^{-3}$ \\
7.14-17.11  & 11.09 & $(3.4^{+1.8}_{-1.3})    \times 10^{-3}$ \\
17-11-29.10 & 22.19 & $(1.1^{+1.4}_{-0.8})    \times 10^{-3}$ \\
29.10-49.09 & 37.44 & $(0.77^{+1.23}_{-0.60}) \times 10^{-3}$ \\
\enddata
\end{deluxetable}

\clearpage

\begin{deluxetable}{cccccccc}
\tabletypesize{\small}
\tablecaption{Antiproton to proton ratio
at the top of the atmosphere
(TOA). \label{t:ratio}}
\tablewidth{0pt}
\tablehead{
Rigidity &\multicolumn{2}{c}{Observed} &
\multicolumn{2}{c}{Extrapolated} & \multicolumn{2}{c}{Atmospheric} &
\\
at & \multicolumn{2}{c}{number of} & \multicolumn{2}{c}
{number\tablenotemark{a}~at top}
&
\multicolumn{2}{c}{secondary} & \( \frac{\mbox{$\overline{{\rm p}}$ }}{{\mbox
p}} \) \\
spectrometer &
\multicolumn{2}{c}{events\tablenotemark{b}} &
\multicolumn{2}{c}{of payload} & & &
 at TOA \\
GV & \mbox{$\overline{{\rm p}}$ } & p & \mbox{$\overline{{\rm p}}$ } & p &
\mbox{$\overline{{\rm p}}$ } & p & }
\startdata
4.0 - 8.0 & 15(1.17) &
85331 & 15.66 & 92201 & 4.06 & 2546 &
$(1.3^{+0.6}_{-0.5}) \times 10^{-4}$ \\
8.0 - 18.0 & 11(0.28) & 39185 & 11.98 & 42417 & 2.64 & 855 &
$(2.3^{+1.2}_{-0.9}) \times 10^{-4}$ \\
18.0 - 30.0 & 3(0.15) & 5765 & 3.16 & 6251 & 0.49 & 113.8 &
$(4.4^{+5.4}_{-3.0}) \times 10^{-4}$ \\
30.0 - 50.0 & 2(0.14) & 1458 & 2.05 & 1583 & 0.14 & 29.0 &
$(1.2^{+2.0}_{-1.0}) \times 10^{-3}$
\enddata
\tablenotetext{a}{The corrections to the top of the payload account only for
loss of particles in the apparatus.}
\tablenotetext{b}{The numbers shown in brackets in column 2 are the
estimated background events due to muons, pions and, in the highest 
energy bin, also spillover protons.}
\end{deluxetable}

\clearpage

\clearpage

\clearpage

\clearpage

\clearpage

\clearpage

\clearpage

\clearpage

\clearpage

\clearpage

\clearpage

\end{document}